\documentclass[aps,prl,twocolumn,showpacs,amsmath,amssymb,nofootinbib,superscriptaddress]{revtex4}
\usepackage{dcolumn}
\usepackage{graphicx}
\usepackage{nicefrac}

\begin{document}

\newcommand{\bra}[1]{\langle #1|}
\newcommand{\ket}[1]{|#1\rangle}
\newcommand{\braket}[2]{\langle #1|#2\rangle}
\newcommand{\reduce}[2]{\langle #1|| \hat{T}(E2) ||#2\rangle}
\newcommand{\reduceE}[2]{\langle #1||M(E2)||#2\rangle}
\newcommand{\element}[2]{$^{#1}$\textrm{#2}}

\title{Early signal of emerging nuclear collectivity in neutron-rich $^{129}$\textrm{Sb} }

\author{T.J.~Gray}
\affiliation{Department of Nuclear Physics, Australian National University, Canberra ACT 0200, Australia}
\author{J.M.~Allmond}
\affiliation{Physics Division, Oak Ridge National Laboratory, Oak Ridge, Tennessee 37831, USA}
\author{A.E.~Stuchbery}
\affiliation{Department of Nuclear Physics, Australian National University, Canberra ACT 0200, Australia}
\author{C.-H.~Yu}
\affiliation{Physics Division, Oak Ridge National Laboratory, Oak Ridge, Tennessee 37831, USA}
\author{C.~Baktash }
\affiliation{Physics Division, Oak Ridge National Laboratory, Oak Ridge, Tennessee 37831, USA}
\author{A.~Gargano}
\affiliation{Istituto Nazionale di Fisica Nucleare, Complesso Universitario di Monte S. Angelo, Via Cintia, I-80126 Napoli, Italy}
\author{A.~Galindo-Uribarri}
\affiliation{Physics Division, Oak Ridge National Laboratory, Oak Ridge, Tennessee 37831, USA}
\affiliation{Department of Physics and Astronomy, University of Tennessee, Knoxville, Tennessee 37996, USA}
\author{D.C.~Radford}
\affiliation{Physics Division, Oak Ridge National Laboratory, Oak Ridge, Tennessee 37831, USA}
\author{J.C.~Batchelder }
\affiliation{Department of Nuclear Engineering, University of California, Berkeley, Berkeley CA 94720, USA}
\author{J.R.~Beene }
\affiliation{Physics Division, Oak Ridge National Laboratory, Oak Ridge, Tennessee 37831, USA}
\author{C.R.~Bingham}
\affiliation{Physics Division, Oak Ridge National Laboratory, Oak Ridge, Tennessee 37831, USA}
\affiliation{Department of Physics and Astronomy, University of Tennessee, Knoxville, Tennessee 37996, USA}
\author{L.~Coraggio}
\affiliation{Istituto Nazionale di Fisica Nucleare, Complesso Universitario di Monte S. Angelo, Via Cintia, I-80126 Napoli, Italy}
\author{A.~Covello}
\affiliation{Dipartimento di Fisica ``Ettore Pancini'', Universit\`{a} di Napoli Federico II, Complesso Universitario di Monte S. Angelo, Via Cintia, I-80126 Napoli, Italy}
\author{M.~Danchev}
\affiliation{Faculty of Physics, St.~Kliment Ohridski University of Sofia, 1164 Sofia, Bulgaria}
\affiliation{Department of Physics and Astronomy, University of Tennessee, Knoxville, Tennessee 37996, USA}
\author{C.J.~Gross }
\affiliation{Physics Division, Oak Ridge National Laboratory, Oak Ridge, Tennessee 37831, USA}
\author{P.A.~Hausladen}
\affiliation{Joint Institute for Heavy Ion Research, Oak Ridge National Laboratory, Oak Ridge, Tennessee 37831, USA}
\author{N.~Itaco}
\affiliation{Istituto Nazionale di Fisica Nucleare, Complesso Universitario di Monte S. Angelo, Via Cintia, I-80126 Napoli, Italy}
\affiliation{Dipartimento di Matematica e Fisica, Universit\`{a} degli Studi della Campania ``Luigi Vanvitelli'', Viale Abramo Lincoln 5, I-81100 Caserta, Italy}
\author{W.~Krolas}
\affiliation{Joint Institute for Heavy Ion Research, Oak Ridge National Laboratory, Oak Ridge, Tennessee 37831, USA}
\affiliation{Institute of Nuclear Physics, Polish Academy of Sciences (IFJ PAN), PL-31342 Krak\'ow, Poland}
\author{J.F.~Liang}
\affiliation{Physics Division, Oak Ridge National Laboratory, Oak Ridge, Tennessee 37831, USA}
\author{E.~Padilla-Rodal }
\affiliation{Joint Institute for Heavy Ion Research, Oak Ridge National Laboratory, Oak Ridge, Tennessee 37831, USA}
\affiliation{Instituto de Ciencias Nucleares, UNAM, AP 70-543, 04510 Mexico City, Mexico}
\author{J.~Pavan}
\affiliation{Joint Institute for Heavy Ion Research, Oak Ridge National Laboratory, Oak Ridge, Tennessee 37831, USA}
\author{D.W.~Stracener}
\affiliation{Physics Division, Oak Ridge National Laboratory, Oak Ridge, Tennessee 37831, USA}
\author{R.L.~Varner}
\affiliation{Physics Division, Oak Ridge National Laboratory, Oak Ridge, Tennessee 37831, USA}

\date{\today}

\begin{abstract}
Radioactive $^{129}$\textrm{Sb}, which can be treated as a proton plus semi-magic $^{128}$\textrm{Sn} core within the particle-core coupling scheme, was studied by Coulomb excitation. Reduced electric quadrupole transition probabilities, $B(E2)$, for the $2^+ \otimes \pi g_{\nicefrac{7}{2}}$ multiplet members and candidate $\pi d_{\nicefrac{5}{2}}$ state were measured. The results indicate that the total electric quadrupole strength of $^{129}$\textrm{Sb} is a factor of 1.39(11) larger than the $^{128}$\textrm{Sn} core, which is in stark contrast to the expectations of the empirically successful particle-core coupling scheme. Shell-model calculations performed with two different sets of nucleon-nucleon interactions suggest that this enhanced collectivity is due to constructive quadrupole coherence in the wavefunctions stemming from the proton-neutron residual interactions, where adding one nucleon to a core near a double-shell closure can have a pronounced effect. The enhanced electric quadrupole strength is an early signal of the emerging nuclear collectivity that becomes dominant away from the shell closure.
\end{abstract}

\pacs{25.70.De, 23.20.-g, 21.10.Ky}

\maketitle

\paragraph*{}%
Atomic nuclei are finite many-body quantum systems that exhibit a unique level of organization. Understanding this organization and the collective phenomena that emerge from the many individual nucleon-nucleon interactions is a leading challenge. The conventional microscopic modeling principle is to first invoke a mean field in which the nucleons move, which establishes the nuclear shell structure, and second, introduce residual interactions between the nucleons outside of a double-shell closure, which leads to configuration mixing and correlations in the nucleonic motion.
\paragraph*{}%
It has long been postulated \cite{SHAGOLD,TALMI} that nuclear collective excitations develop when the long-range part of the proton-neutron (pn) residual interaction, which is thought to drive the emergence of collectivity and deformation, overcomes the short-range pairing interaction, which akin to Cooper-pair formation in superconductors couples like nucleon pairs to spin zero and favors spherical shapes. The long-range pn interaction increases as both protons and neutrons are added outside a closed shell. Thus, the quest to understand how collectivity emerges, and the role of proton-neutron interactions, is traditionally based on systematic studies of sequences of nuclei that exhibit increasing collectivity, starting at a closed shell.
\paragraph*{}%
One of the simplest possible steps that can be taken is to study the change in collectivity accompanying the addition of a single nucleon outside of a semi-magic even-even core. Nuclear collectivity is signalled by strong electric quadrupole ($E2$) transitions between low-excitation energy levels. In seeking to understand the emergence of nuclear collectivity, it is essential to study $E2$ transition strengths, which may begin to show collective features before the patterns associated with deformed collective excitations (e.g., anharmonic vibrations and rotations) emerge in the energy levels.
\paragraph*{}%
The region around double-magic \element{132}{Sn} is now accessible through experiments on radioactive beams. This provides an excellent opportunity to investigate the emergence of nuclear collectivity from the underlying single-particle motion because \element{132}{Sn} is a robust doubly magic core~\cite{RAD,KATE,JMA133,ROS,GORG}. In particular, \element{132}{Sn} does not have deformed multi-particle, multi-hole states at low-excitation energy (like \element{16}{O} and \element{40}{Ca}) that can mix with the lowest-lying states and complicate the interpretation of shell-model calculations. 
\paragraph*{}%
The framework for our present investigation into the emergence of collectivity near  \element{132}{Sn} is the particle-core coupling concept introduced by de-Shalit \cite{DESHAL}, further developed by Bohr and Mottleson~\cite{BM52,BM53,BM60,BM75}, and used in modern effective field theory calculations \cite{COEL}. In this model, a single nucleon in an orbit of angular momentum $j$ is coupled to the 0$^+$ ground state and the first $2^+$ excitation of an even-even core. The odd-mass nuclide has a ground-state angular momentum of $j$ and a ``multiplet" of states near the excitation energy of the core 2$^+$ state with angular momentum $I$, where $|j-2| \leqslant I \leqslant | j+2|$, formed by coupling the odd nucleon to the 2$^+$ core excitation. The assumption that the odd nucleon does not perturb the core, together with angular momentum coupling algebra, gives rise to an $E2$ sum rule: $\sum  B(E2;\uparrow)_{\mathrm{multiplet}} = B(E2; 0^+ \rightarrow 2^+)_{\mathrm{core}}$. 
This sum rule, which is implicit in particle-core coupling models and used in textbook examples of collective structure in odd nuclei \cite{RW}, was empirically demonstrated in 1976 by Tuttle et al. \cite{TUTT}. The seminal studies on \element{113,115}{In} (a proton hole in $Z=50$) \cite{TUTT,DIET} have revealed total electric quadrupole strengths that are consistent with those of their \element{114,116}{Sn} cores at the neutron midshell. While the required data remain scarce, the sum rule has been empirically robust to date (as we will demonstrate below).
\paragraph*{}%
In this Letter we report rare evidence of the break-down of the particle-core $E2$ sum rule. Importantly, the $E2$ strength observed in $^{129}_{\;\: 51}$Sb$_{78}$ {\em exceeds} that of its $^{128}_{\;\: 50}$Sn$_{78}$ core, indicating an enhancement of collectivity as a result of the added proton. Shell-model calculations show that the enhanced collectivity originates from coherent contributions of the valence proton and neutrons together, which can be interpreted as an early indication of emerging collectivity in a nuclear system only four neutron holes and one proton away from doubly magic \element{132}{Sn}.

\paragraph*{}
A radioactive ion beam of \element{129}{Sb} at an energy of 400~MeV was Coulomb excited on a 1.0-mg/cm$^2$ self-supporting enriched \element{50}{Ti} target. The measurement was performed at the Holifield Radioactive Ion Beam Facility (HRIBF) of Oak Ridge National Laboratory (ORNL). Recoiling target-nuclei were measured in a 2$\pi$ CsI array, BareBall \cite{BAREBALL}, and subsequent $\gamma$ rays were measured in a Compton-suppressed HPGe Clover array, CLARION \cite{GROSS}. A Bragg-Curve gas detector was used to measure beam compositions and stopping powers. $B(E2)$ values were determined by measuring cross sections and particle-$\gamma$ angular correlations of excited states following Coulomb excitation.

\paragraph*{}
The isotopic composition of the target was subsequently measured by inductively coupled plasma mass spectrometry, giving 1.64(3)$\%$ \element{46}{Ti}, 1.35(3)$\%$  \element{47}{Ti}, 12.09(12)$\%$  \element{48}{Ti}, 3.52(4)$\%$ \element{49}{Ti}, and 81.40(81)$\%$ \element{50}{Ti}. The beam composition was directly measured with a zero-degree Bragg detector, resulting in 6.2(7)$\%$ \element{129}{Sn}, 41(2)$\%$ \element{129}{Sb}, and 52(1)$\%$ \element{129}{Te}. A preliminary spectrum of the beam composition was given in Ref.~\cite{CHYU}. The ground and isomeric components of the beam were measured by decay at the center of CLARION. For the \element{129}{Sb} beam component of interest, 63(2)$\%$ was in the ground state and 37(2)$\%$ was in the isomeric $\nicefrac{19}{2}^-$ state. The energy loss of the beam through the target was determined to be 56(2)~MeV from the Bragg detector.

\begin{figure}[]
\centering
\includegraphics[width=3.35in]{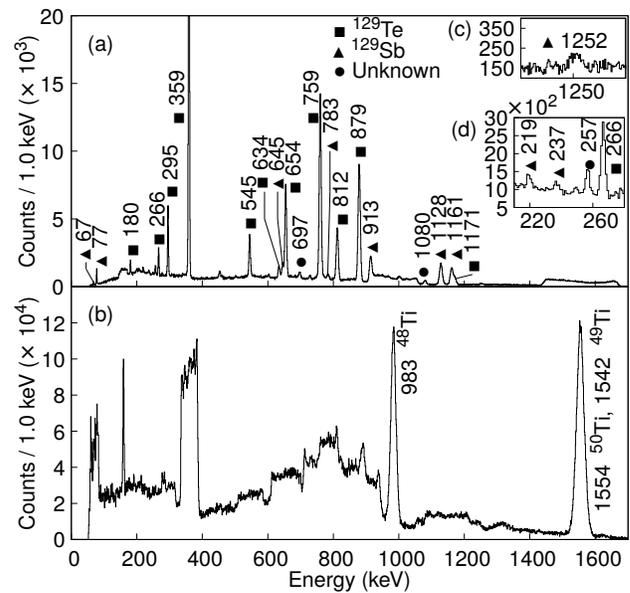}
\caption{The Ti-gated $\gamma$-ray spectra Doppler corrected for (a) $A=129$ beam, and (b) Ti target. Insets (c) and (d) show some of the weaker features of the $A=129$ transitions.}
\label{fig:e_spectra}
\end{figure}

\begin{figure}[]
\centering
\includegraphics[width=3.0in]{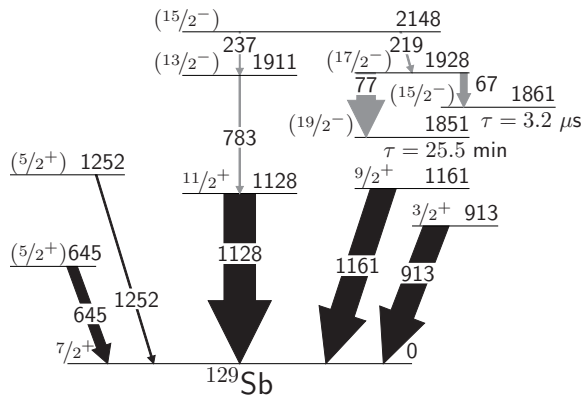}
\caption{Partial level scheme for \element{129}{Sb}. Grey transitions relate to excitation of the 1851-keV, 25.5-min isomer present in the beam and subsequent decay. This part of the level scheme is not drawn to scale.}
\label{fig:Sb129}
\end{figure}

\paragraph*{}
The \textrm{Ti}-gated $\gamma$-ray spectra are shown in Fig.~\ref{fig:e_spectra} and a partial level scheme for the states and transitions observed in \element{129}{Sb} is given in Fig.~\ref{fig:Sb129}. Many of the lines in Fig.~\ref{fig:e_spectra}(a) are associated with known lines in \element{129}{Te}. Turning to \element{129}{Sb}, the most strongly excited states are the four low-lying states directly connected to the ground state at 645, 913, 1161 and 1128 keV. There is also small but clear two-step Coulomb excitation of the $\nicefrac{19}{2}^-$ isomeric state present in the beam. Due to the lack of a good efficiency calibration at low energy, Coulomb-excitation analysis in the present work is limited to ground-state excitation. Three relatively weak unidentified transitions at 257, 697, and 1080~keV in the particle-$\gamma$ spectrum were disregarded for the Coulomb-excitation analysis; none of these transitions are observed in $\gamma$-$\gamma$ coincidences. Figure~\ref{fig:e_spectra}(b) shows the $\gamma$-ray spectrum Doppler corrected for the \textrm{Ti} target.

\paragraph*{}
Based on the weak particle-core coupling limit \cite{DESHAL}, where the excitation strength scales with $(2I + 1) / (2\times\nicefrac{7}{2} + 1)$ from the core strength, the two strongest transitions to the $\nicefrac{7}{2}^+$ ground state are expected to be from the $\nicefrac{11}{2}^+$ (parallel) and $\nicefrac{9}{2}^+$ multiplet members. However, shell-model calculations, which will be introduced later, predict the third strongest transition to be from the $\nicefrac{3}{2}^+$ (anti parallel) multiplet member and relatively weak transitions from the $\nicefrac{5}{2}^+$ and $\nicefrac{7}{2}^+$ multiplet members. The three strong transitions observed are from states at 913, 1128, and 1161 keV, see Fig.~\ref{fig:Sb129}. The 913-keV state is assigned $I^\pi = \nicefrac{3}{2}^+$ from ($^{3}$H,$\alpha$) transfer reactions which show that the angular momentum transfer $L(^{3}{\mathrm H},\alpha)=2$~\cite{ENTWISLE, CONJEAUD}. The 1161-keV state can be identified as $\nicefrac{9}{2}^+$ because it is populated indirectly in the $\beta^-$ decay from the \element{129}{Sn} ground state ($I^\pi = \nicefrac{3}{2}^+$) \cite{STONE}, whereas the 1128-keV state is not. This leaves the 1128-keV state, which is populated in various isomer and prompt-fission decays~\cite{ENSDF}, as the only candidate for $\nicefrac{11}{2}^+$. These three assignments agree with Ref.~\cite{ENSDF}. The 1252-keV state is tentatively assigned $I^\pi = \nicefrac{5}{2}^+$ on the basis of energy systematics of the \textrm{Sb} isotopes \cite{ENSDF} and direct population from single-step Coulomb excitation in the present study. The $(2I+1)$ weighted mean energy of the tightly grouped set of states between 913 and 1252 keV is 1135 keV \cite{ENSDF}, which is in agreement with the first $2^+$ energy of the \element{128}{Sn} core, 1168 keV, suggesting minimal contribution from components other than $\nu2^+ \otimes \pi g_{\nicefrac{7}{2}}$.

\paragraph*{}
The Coulomb-excitation analysis was performed using the semi-classical program \textsc{gosia} \cite{GOSIA}. The analysis procedures, including necessary corrections, were similar to those in Refs.~\cite{JMA, JMASN2, STUCHTE, JMANI, JMASN,JMATE136}. The absolute $B(E2)$ values were extracted relative to \element{48}{Ti} with $B(E2; 0_1^+\rightarrow 2_1^+)=0.0662(29)~\text{e}^2\text{b}^2$ \cite{BORIS}. Uncertainties due to unknown branching ratios, $\delta=\nicefrac{E2}{M1}$ mixing ratios, quadrupole moments, and interference effects were included in the analysis; experimental limits were used where possible. The ``safe'' criterion, cf.~\cite{CLI,LES,JMASAFE,JMATE136}, was investigated with the reaction program \textsc{fresco} \cite{FRESCO}. For the nuclear potentials set to zero, the \textsc{fresco} calculations agreed to within $3.6\%$ of the \textsc{gosia} calculations. For the most backward center-of-mass angles, \textsc{fresco} calculations --- with real and imaginary potentials up to $50$ and $10$~MeV, respectively --- showed destructive Coulomb-nuclear interference effects with up to $12\%$ smaller cross sections. These effects would result in $B(E2)$s that are too small, and it would have less impact on the more forward center-of-mass angles measured. Overall, the extracted $B(E2)$ values were consistent as a function of center-of-mass angle within statistical uncertainty.

\begin{figure}[]
\centering
\includegraphics[width=3.35in]{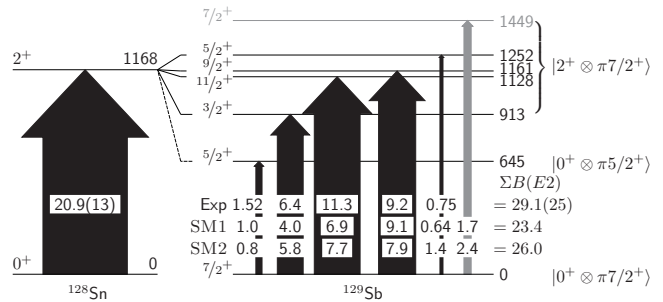}
\caption{The fragmentation of $E2$ strength in W.u. over the $2^+ \otimes \pi g_{\nicefrac{7}{2}}$ multiplet members and candidate $\pi d_{\nicefrac{5}{2}}$ state of \element{129}{Sb} and enhancement of total strength as compared to the \element{128}{Sn} core \cite{JMA} are shown. For comparison, the $B(E2;0_1^+\rightarrow2_1^+)$ for \element{130}{Te} is 74.4(26)~W.u. \cite{STUCHTE}. The grey colored transition was not experimentally observed.}
\label{fig:level_schemes}
\end{figure}

\paragraph*{}
The excitation $B(E2)$ values and fragmentation of strength over the $2^+ \otimes \pi g_{\nicefrac{7}{2}}$ multiplet members and candidate $\pi d_{\nicefrac{5}{2}}$ state are shown in Fig.~\ref{fig:level_schemes}. The primary observation is that the fragmented $B(E2)$ strength sums to a value that is a factor of 1.39(11) larger than the \element{128}{Sn} core excitation \cite{JMA}. This is in stark contrast to the expectation of equal sums in particle-core coupling schemes, which do not modify the core or develop extra total collectivity due to particle-core interactions. The spectroscopic results are compared to two shell-model calculations in Table~\ref{tab:ShellModelComp} and the calculated $B(E2)$ values and sums are also shown in Fig.~\ref{fig:level_schemes}. Within the general weak particle-core coupling limit \cite{DESHAL}, the $B(E2;\downarrow)$ of each multiplet member should be equal to each other, which is clearly not the case; the results are qualitatively more consistent with intermediate coupling.

\begingroup
\squeezetable
\begin{table*}[]
  \centering
  \caption{$B(E2)$ results of \element{129}{Sb} and comparison to shell-model calculations.}
  \label{tab:ShellModelComp}
\begin{ruledtabular}
  \begin{tabular}{llllllllll}
    & \multicolumn{3}{c}{Experiment} & \multicolumn{3}{c}{SM1} & \multicolumn{3}{c}{SM2} \\ \cline{2-4} \cline{5-7} \cline{8-10}
    $I_i^\pi$ & $E_x$ & $B(E2\uparrow)$  & $B(E2\downarrow)$ & $E_x$ & $B(E2\uparrow)$ & $B(E2\downarrow)$ & $E_x$ & $B(E2\uparrow)$ & $B(E2\downarrow)$ \\
    & (keV) & (W.u.)  & (W.u.) &  (keV) & (W.u.) & (W.u.) & (keV) & (W.u.) & (W.u.) \\ \hline
    $\nicefrac{5}{2}_1^+$ 	& 645 	& 1.52(25)	& 2.0(3) 	& 937   & 1.0 	& 1.4 & 781   & 0.80 	& 1.1 \\
    $\nicefrac{3}{2}_1^+$ 	& 913 	& 6.4(7) 	& 12.7(14)& 1090 & 4.0 	& 8.0 & 1204 & 5.8 	& 11.6 \\
    $\nicefrac{11}{2}_1^+$ 	& 1128 	& 11.3(7) 	& 7.5(5) 	& 1172 & 6.9 	& 4.6 & 1419 & 7.7 	& 5.1 \\
    $\nicefrac{9}{2}_1^+$ 	& 1161 	& 9.2(8) 	& 7.3(6) 	& 1078 & 9.1 	& 7.3 & 1417 & 7.9 	& 6.3 \\
    $\nicefrac{5}{2}_2^+$ 	& 1252 	& 0.75(9) 	& 1.0(12) 	& 1245 & 0.64 	& 0.85 & 1440 & 1.4	& 1.9 \\
    $\nicefrac{7}{2}_2^+$ 	&     		&          	&         	& 1449 & 1.7 	& 1.7 & 1695 & 2.4 	& 2.4 \\ \cline{1-1} \cline{3-3} \cline{6-6} \cline{9-9}
    $\sum B(E2\uparrow)$ 	&     		& 29.1(25)&         	&      	    & 23.4 	&    	 &          & 26.0  & 	\\
\end{tabular}
\end{ruledtabular}
\end{table*}
\endgroup

\paragraph*{}
Shell-model calculations were performed with the \textsc{NuShellX} \cite{NUSH} (SM1) and \textsc{antoine} \cite{ANT} (SM2) programs using different nucleon-nucleon interactions as described in Refs.~\cite{Brown2005} and \cite{Coraggio2017}, respectively. These two independent calculations represent state-of-the-art shell-model calculations near $^{132}$Sn. Both calculations use a \element{100}{Sn} core and include the $0g_{\nicefrac{7}{2}}$,  $1d_{\nicefrac{5}{2}}$, $1d_{\nicefrac{3}{2}}$, $2s_{\nicefrac{1}{2}}$, and $0h_{\nicefrac{11}{2}}$ orbitals for protons and neutrons. Both interactions are based on the CD-Bonn nucleon-nucleon potential, but with different procedures for renormalization and derivation of the effective Hamiltonian; both add a Coulomb term for the proton-proton interaction. The interaction used with SM1 is designated jj55pn. The SM1 calculations used an effective proton charge of $e_p = 1.7e$, and an effective neutron charge of $e_n = 0.8e$ chosen to reproduce the experimental $B(E2;0_1^+\rightarrow2_1^+)$ values in $^{134}_{\;\: 52}$Te$_{82}$ and $^{128}_{\;\: 50}$Sn$_{78}$~\cite{STUCHTE,JMA}. The SM2 calculations used $e_p = 1.7e$ and $e_n = 0.9e$, which also reproduce these closed-shell $B(E2)$ values. 

\paragraph*{}
The shell-model calculations account for much of the enhanced collectivity. Both calculations give a low-lying $\nicefrac{5}{2}^+$ state that is dominated by a $d_{\nicefrac{5}{2}}$ proton, and five states with $I^\pi = \nicefrac{3}{2}^+,\nicefrac{5}{2}^+,\nicefrac{7}{2}^+,\nicefrac{9}{2}^+,\nicefrac{11}{2}^+$, which  correspond to the $\nu 2^+ \otimes \pi g_{\nicefrac{7}{2}}$ multiplet. The energy of the $\nicefrac{5}{2}_1^+$ state is higher than in experiment in both calculations, with SM2 giving a better result. However, SM2 gives multiplet energies that are $\sim$200-keV too high whilst the SM1 multiplet energies are closer to experiment.

\paragraph*{}
The $B(E2)$ strengths are under-predicted in both calculations, especially for the parallel and anti-parallel spin-coupled $\nicefrac{11}{2}^+$ and $\nicefrac{3}{2}^+$ states, with SM2 giving higher $B(E2)$ strengths than SM1. If the nominal effective charges of SM2 are used for SM1, the $B(E2\uparrow)$ values increase with a new total sum of 28.6 W.u.\ but it comes at the expense of over predicting the \element{128}{Sn} $B(E2\uparrow)$; the ratio of summed strengths remains unchanged but it can be made larger by small changes to the harmonic oscillator wave functions and $\hbar\omega$ scaling. Considering the differences between SM1 and SM2, uncertainties in the effective charges, and sensitivity to the harmonic oscillator parameters, the short fall in $E2$ strength in the SM calculations is on the order of the theoretical uncertainties. The following discussion focuses on the origin of the enhanced collectivity, which is at least qualitatively predicted by both SM1 and SM2.

\paragraph*{}
The wave functions generated by the two shell-model calculations predict the same dominant configuration in each state but the SM2 wavefunctions are more fragmented over the configuration space; the predicted larger values of the $B(E2)$ transition strengths in SM2 must therefore be associated with constructive interference among the many contributions to the $B(E2)$ values, which can be interpreted as an indication of emerging collectivity.

\paragraph*{}
According to both shell-model calculations, the odd proton in \element{129}{Sb} changes the configuration mixture in the neutron part of the wavefunction compared to the \element{128}{Sn} core. The consequences can be investigated by examining the proton and neutron components of the $B(E2)$:
$B(E2; I_i \rightarrow I_f) = (e_p A_p + e_n A_n)^2/(2 I_i + 1) = (e_p^2 A_p^2 + e_n^2 A_n^2 +2e_p e_n A_p A_n)/(2 I_i + 1)$,
where $A_p$($A_n$) is the reduced matrix element for the proton(neutron) contribution to the transition. A first observation is that the addition of the proton {\em does not increase} the collectivity of the $^{128}$Sn core - the $A_n^2$ term alone in the sum rule only just matches that of the core in SM2 and falls short by about 10\% for SM1. The $A_p^2$ term due to the single proton contributes less than 10\% to the sum rule in both calculations. Thus in both calculations it is the pn term, $e_p e_n A_p A_n$, adding coherently overall to the $A_p^2$ and $A_n^2$ terms, that gives the additional strength in the sum. This is a clear signature that the residual pn interactions are not only fragmenting the wavefunctions, but doing so in a way that leads to constructive interference in the $E2$ strength. Thus the proton-neutron interactions, which cause the splitting of the $2^+ \otimes \pi g_{7/2}$ multiplet, also cause the increased collectivity. A recent study  \cite{PETERS} of emerging collectivity in \element{132}{Xe}, the isotone of \element{129}{Sb}, demonstrated similar fragmentation of the wavefunctions together with emerging quadrupole correlations, as neutron pairs are removed from \element{136}{Xe} ($N=82$). The present study on \element{129}{Sb} with a single valence proton exposes the role of the proton-neutron interactions in the emergence of collectivity.

\begin{figure}[]
\centering
\includegraphics[width=2.6in]{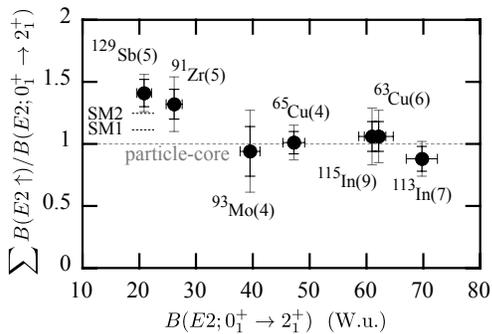}
\caption{A global view of particle-core to core $\sum B(E2\uparrow)$ ratios for nuclei adjacent to shell closures. The number of transitions included in $\sum B(E2\uparrow)$ for the odd-mass nuclei are given in parenthesis and the outer error bar reflects a linear sum of uncertainties for $\sum B(E2\uparrow)$, which is more appropriate if the transitions are dominated by shared systematic uncertainties. The data are from Refs.~\cite{TUTT,ENSDF, BORIS, JMA, JMANI, JMASN}.}
\label{fig:SumPlot}
\end{figure}

\paragraph*{}
A global comparison of electric quadrupole strength, $\sum_i B(E2; I^\pi_{\rm g.s.} \rightarrow I^\pi_i)$, between semi-magic even-even and adjacent odd-mass nuclei is shown in Fig.~\ref{fig:SumPlot}, expressed as ratios. For the semi-magic nuclei this sum is simply $B(E2; 0_1^+ \rightarrow 2_1^+)$; for the odd-mass nuclei, it is the sum of all $B(E2\uparrow)$ values connected to the ground state.  \element{129}{Sb}, and to a lesser extent \element{91}{Zr}, deviate from the simple expectation that the odd-mass nucleus and semi-magic core should have equal $B(E2\uparrow)$ values. The \element{129}{Sb} result provides a clear indication of an experimental departure from the hitherto empirically successful sum rule. 
Adding one nucleon to a core near a double-shell closure with a small $B(E2)$ value can have a pronounced effect giving sensitivity to emerging collectivity that otherwise would be obscured.

\paragraph*{}
In conclusion, radioactive $^{129}$\textrm{Sb}, which can be considered as a proton plus semi-magic $^{128}$\textrm{Sn} core within the particle-core coupling scheme, was studied by Coulomb excitation. The reduced electric quadrupole transition probabilities, $B(E2)$, for the $2^+ \otimes \pi g_{\nicefrac{7}{2}}$ multiplet members and candidate $\pi d_{\nicefrac{5}{2}}$ state were measured. The results indicate that the total electric quadrupole strength of $^{129}$\textrm{Sb} is a factor of 1.39(11) larger than that of the $^{128}$\textrm{Sn} core, providing an early signal of the emerging nuclear collectivity that becomes dominant away from shell closures. Shell-model calculations suggest that the enhancement is due to increased sensitivity (i.e., when in proximity to a double-shell closure) to constructive quadrupole coherence stemming from proton-neutron interactions. The present study on \element{129}{Sb} with a single valence proton explicitly exposes and quantifies the role of the proton-neutron interaction in the emergence of collectivity.

\begin{acknowledgments}
The authors gratefully acknowledge the HRIBF operations staff for providing the beams used in this study and T.~Papenbrock, A.~Volya, and J.L.~Wood for fruitful discussions. This material is based upon work supported by the U.S. Department of Energy, Office of Science, Office of Nuclear Physics, under Contract No.\ DE-AC05-00OR22725, and this research used resources of the Holifield Radioactive Ion Beam Facility of Oak Ridge National Laboratory, which was a DOE Office of Science User Facility. This research was also sponsored by the Australian Research Council under grants No.\ DP0773273 and DP170101673, and by the U.S.\ DOE under Contract No.\ DE-FG02-96ER40963 (UTK). T.J.G. acknowledges the support of the Australian Government Research Training Program. E.P-R. acknowledges the financial support of PAPIIT-IN110418.
\end{acknowledgments}

\end{document}